\documentclass[twocolumn,english]{IEEEtran}
\usepackage[T1]{fontenc}
\usepackage[latin9]{inputenc}
\usepackage{xcolor}
\usepackage{amsmath}
\usepackage{amsthm}
\usepackage{graphicx}
\usepackage{graphics}
\usepackage{caption}
\usepackage{subcaption}
\usepackage{stmaryrd}
\usepackage{amssymb}
\usepackage{dirtytalk}
\usepackage{algorithm,algorithmicx}
\usepackage[noend]{algpseudocode}
\usepackage{tabularx}
\usepackage{wrapfig}
\usepackage{enumitem}
\usepackage{mathtools}
\usepackage{multirow}
\usepackage{float}
\usepackage{hyperref}
\usepackage[colorinlistoftodos]{todonotes}

\makeatletter

\theoremstyle{plain}
\newtheorem{thm}{\protect\theoremname}
\theoremstyle{definition}
\newtheorem{defn}[thm]{\protect\definitionname}
\newcommand\ddfrac[2]{{\displaystyle\frac{\displaystyle #1}{\displaystyle #2}}}

\algnewcommand\algorithmicforeach{\textbf{for each}}
\algdef{S}[FOR]{ForEach}[1]{\algorithmicforeach\ #1\ \algorithmicdo}

\newcommand{\node}[1]{\text{\texttt{#1}}}

\usepackage{babel}

\newcolumntype{L}[1]{>{\raggedright\arraybackslash}p{#1}}
\newcolumntype{C}[1]{>{\centering\arraybackslash}p{#1}}

\providecommand{\definitionname}{Definition}
\providecommand{\theoremname}{Theorem}

\DeclareRobustCommand\widecheck[1]{{\mathpalette\@widecheck{#1}}}
\def\@widecheck#1#2{%
    \setbox\z@\hbox{\m@th$#1#2$}%
    \setbox\tw@\hbox{\m@th$#1%
       \widehat{%
          \vrule\@width\z@\@height\ht\z@
          \vrule\@height\z@\@width\wd\z@}$}%
    \dp\tw@-\ht\z@
    \@tempdima\ht\z@ \advance\@tempdima2\ht\tw@ \divide\@tempdima\thr@@
    \setbox\tw@\hbox{%
       \raise\@tempdima\hbox{\scalebox{1}[-1]{\lower\@tempdima\box
\tw@}}}%
    {\ooalign{\box\tw@ \cr \box\z@}}}

\makeatother

\usepackage{babel}
\providecommand{\definitionname}{Definition}
\providecommand{\theoremname}{Theorem}

\begin{document}
\global\long\def\var{\mathrm{var}}%

\global\long\def\ER{Erd\"{o}s-R\'{e}nyi}%

\global\long\def\BA{Barab\'{a}si-Albert}%

\title{Graph Compression with Application to Model Selection}

\author{Mojtaba Abolfazli, Anders H{\o}st-Madsen, June Zhang, Andras Bratincsak\thanks{M. Abolfazli, A. H{\o}st-Madsen, and J. Zhang are with the Department of Electrical and Computer Engineering, University of Hawaii at Manoa, 2540 Dole Street, Honolulu, HI 96822; e-mail: \{mojtaba,ahm,zjz\}@hawaii.edu. 

A. Bratincsak is with the Department of Pediatrics, John A. Burns School of Medicine, University of Hawaii, Honolulu, HI 96813; e-mail: andrasb@hphmg.org. 

The research was funded in part by the NSF grant CCF-1908957.}}

\maketitle
\begin{abstract}
Many multivariate data such as social and biological data exhibit complex dependencies that are best characterized by graphs. Unlike sequential data, graphs are, in general, unordered structures. This means we can no longer use classic, sequential-based compression methods on these graph-based data.
Therefore, it is necessary to develop new methods for graph compression. In this paper, we present universal source coding methods for the lossless compression of unweighted, undirected, unlabelled graphs. We encode in two steps: 1) transforming graph into a rooted binary tree, 2) the encoding rooted binary tree using graph statistics. Our coders showed better compression performance than other source coding methods on both synthetic and real-world graphs. 

We then applied our graph coding methods for model selection of Gaussian graphical models using minimum description length (MDL) principle finding the description length of the conditional independence graph. Experiments on synthetic data show that our approach gives better performance compared to common model selection methods. We also applied our approach to electrocardiogram (ECG) data in order to explore the differences between graph models of two groups of subjects. 

\end{abstract}

\section{Introduction}
Graphs can be used to represent complex dependencies between many variables. Graph-based analysis are used in many disciplines such as social networks analysis, natural language processing, chemistry, and bioinformatics. Real-world graphs can be large and expensive to store and communicate. Developing efficient methods to compress graphical data are of interest for storage and transmission of data.
Classic information theory methods for coding deal with sequential data. Unlike sequential data with starting and stopping points (ordering), graphs are unordered structures. 
Since  ordering is essential in coding, graph coding is challenging. This means that graphs cannot be efficiently compressed using traditional, sequence-based compression methods.

In this paper, we first present new universal source coding methods for the lossless compression of unlabeled, unweighted, undirected, simple graphs. Since the graphs are unlabeled, vertices do not have distinct identifications except through their interconnections. Reference \cite{ChoiSzpankowski12} referred to such a graph as a graph \emph{structure}. 
Our approach has two steps. First, inspired by Steinruecken's method for coding of unordered i.i.d sequences \cite{Steinruecken15}, we transform a graph, $G$, into an equivalent rooted binary tree, $T$. 
Then, we use graph statistics from $G$ to develop two classes of coders for $T$. The first class utilizes local properties (i.e., formation of graph motifs) such as triangles and the second class uses degree distribution, as a global graph statistics, along with local properties for encoding. This step uses past information (encoded vertices) to encode the connections at the current vertex. 
This way we can build a probabilistic model based on graph statistics (either by learning or estimating statistics) to encode graphs. Our coders reflect more information about the structure of $G$ and therefore give shorter 
codelengths compared to structural coding of \cite{ChoiSzpankowski12} for
graphs with more structure than
\ER\ graphs.

In the second half of the paper, we use graph coding for data analysis by using
description length, which is the number of bits to describe data. Rissanen proposed using description length for model selection in his work on the minimum description length (MDL)  \cite{Rissanen83, Rissanen78, Rissanen86}. Rissanen's MDL principle codifies model selection, which balances between how well a model describes the given data and the complexity of the model. Here, we develop a novel approach for model selection of Gaussian graphical models. We compute the description length of the conditional independence graph (the sparsity pattern of the precision matrix) using our lossless graph coding methods and the data under the assumed Gaussian distribution. The model that minimizes the summation of these two terms will be selected as the best model. Unlike other methods that may only consider the number of edges in conditional independence graph to account for model complexity, our approach considers the whole structure of the conditional independence graph by lossless compression.

We showed using synthetic and real-world data the advantages of our methods for both compression and Gaussian graphical model selection. 
Our approach outperforms competing source coding methods in two different scenarios: 1) compression of a single real-world graph, 2) compression of a graph from a particular type after learning the statistics of its type through training.
We also compared our approach with common methods in the literature for model selection in Gaussian graphical models. The experimental results on synthetic data showed that our approach can recover true graph model of data with higher F1-score. We also considered a real-world dataset containing extracted features from 12-lead electrocardiogram (ECG) signals of a group of healthy people and a group of people with Kawasaki disease. We observed that there is a difference between graph model of healthy people and those with Kawasaki disease.

The paper is organized as follows. First, we provide an overview of previous works on universal compression methods for graphs in Section~\ref{PriorWork.sec}.
In Section~\ref{Coding.sec}, we describe our approach for transforming a graph structure into a rooted binary tree and show that the compression of rooted binary tree is same as the compression of graph structure. Then, we introduce two classes of universal coders based on graph statistics and provide experimental results on synthetic and real-world graphs. In Section~\ref{GLASSO.sec}, we present the application of graph coding for model selection in Gaussian graphical models. We also give the performance results on synthetic data and then apply our approach to ECG data. We make concluding remarks and give some directions for future work in Section~\ref{Conclusion.sec}.

\subsection{Prior Work \label{PriorWork.sec}}
Graph compression is a relatively new area in source coding. References \cite{LuczakSzpankowski17,LuczakSzpankowski17b,AsadiAbbeVerdu17, delgosha2020graphical} focused on the entropy analysis of graph compression. 
Other papers have provided practical graph compression algorithms. These algorithms can be designed for the compression of either unlabeled graph (i.e., graph structure) or labeled graph (i.e., encoding vertices labels together with graph structure). In the case of unlabeled graph, the decoder recovers a graph that is isomorphic to the input graph. 
In the case of labeled graph, the decoder recovers the exact graph (i.e., graph structure with labels) at the expense of longer codewords. In other words, encoding of unlabeled graphs benefits from isomorphism since there are different labeled graphs that have the same structure. Fewer bits are required, in general, to encode an unlabeled graph compared to a labeled graph with the same structure \cite{ChoiSzpankowski12}.

Reference \cite{ChoiSzpankowski12} was the first study to develop a compression algorithm to reach the entropy of the distribution on unlabeled \ER\ graphs up to the first two terms. In \cite{host2018coding}, two universal coding methods for arbitrary unlabeled graphs based on degree distribution and formation of triangles were introduced. 
The authors in \cite{luczak2019compression} presented asymptotically optimal structural compression algorithms for the compression of both unlabeled and labeled preferential attachment graphs. The compression of dynamic graphs generated by duplication model was studied in \cite{turowski2020compression}. The authors developed algorithms for the compression of both unlabeled and labeled versions of such graphs.
Reference \cite{basu2018universal} introduced an algorithm for the compression of deep feedforward neural networks by modeling them as bipartite graph layers. An algorithm for the compression of sparse labeled graphs with marks on vertices and edges was introduced in \cite{delgosha2020universal}. A general survey on lossless graph compression methods can be found in \cite{besta2018survey}.

Previous compression methods are tailored to specific graph models and can perform poorly on other graph models or real-world graphs. Our approach provides a method to encode arbitrary unlabeled graph structures by building a probabilistic model based on graph properties (e.g., graph motifs and degree distribution). This general approach extracts more information from the graph structure than prior work and results in shorter codelength compared to other source coding methods.  

The literature on model selection methods for Gaussian graphical models is rich. Existing methods can broadly fall into two main classes: information-based methods and resampling methods. Information-based methods, such as Bayesian information criteria (BIC)  \cite{yuan2007model}, Akaike information criteria (AIC) \cite{menendez2010gene}, and extended Bayesian information criteria (EBIC) \cite{foygel2010extended}, select the best model based on the log-likelihood of the data and the complexity of Gaussian graphical model.
To account for model complexity, these methods typically consider the number of edges from the conditional independence graph, $G$, induced by the Gaussian graphical model; while easy to obtain, this statistic gives only a rough approximation of the structure of $G$. In contrast, our approach for model selection is to choose the best model based on the description length of the $G$ and the data when encoded with the resulting conditional independence graph \cite{abolfazli2021graph}.  

Resampling methods measure the performance on out-of-sample data by splitting the data into into a subset of samples for fitting and use the remaining samples to estimate the efficacy of the model. The most common method in this class is cross-validation (CV) \cite{rothman2008sparse}. Other methods are Generalized Approximate Cross-validation (GACV) \cite{lian2011shrinkage}, Rotation Information Criterion (RIC) \cite{lysen2009permuted}, and Stability Approach to Regularization Selection (StARS) \cite{liu2010stability}.
The major shortcoming of resampling methods is their high computational cost since they require to solve the problem across all sets of subsamples.

A key aspect of model selection in Gaussian graphical models is the structure of the conditional independence graph. 
Resampling methods overlook this aspect while information-based methods only consider simple graph statistics such as the number of edges in the conditional independence graph. Our approach presents a different perspective to account for a more accurate model complexity. This relies on being able to compute the description length of the conditional independence graph, $G$, using our lossless graph compression methods.

\section{\label{Coding.sec}Graph Compression}
Consider an unweighted, undirected, simple, unlabeled graph $G(V,E)$, where $V$ is the set of vertices and $E$ is the set of edges.
In this paper, we present a method to compress a graph structure inspired by Steinruecken's method for coding unordered i.i.d binary sequences \cite{Steinruecken15}. We use
the terms compression and coding interchangeably in this paper as they denote the same process.
One can see our methods as extension
of \cite{ChoiSzpankowski12} to take into
account more graph structure.
The general idea is to randomly pick a vertex, $V_i \in V$, and encode information about all the incident edges of $V_i$. A neighbor of $V_i$ is then picked and the information about its incident edges are encoded. To implement this scheme, an unlabeled graph $G$ is transformed into a rooted binary tree $T$. Each level of $T$ associated with the vertex $V_i$ from $G$ to be encoded. An encoder is designed to encode $T$. A decoder will decode $T$ without error to recover $\widetilde{G}(V,E)$, which is identical to $G(V,E)$ up to an automorphism of the nodes. In this paper, we introduce two broad classes of graph coders to efficiently encode the rooted binary tree $T$.

\subsection{Definitions and Notation}
Let $V_i$  denote the $i$th vertex in $G$. We will exclusively use \emph{vertices} to refer to the nodes in $G$ and \emph{nodes} to refer to the nodes in $T$. The total number of vertices in G is $|V|$. Let $V_i \leftrightarrow V_j$ denote an edge between vertices $V_i$ and $V_j$ in $G$. The degree of vertex $V_i$ is the total number of vertices connected to $V_i$. The degree distribution $P(k)$ is the probability distribution of degrees of vertices in $G$ and is an often used statistics to differentiate between different classes of random graphs.
The adjacency matrix of $G$ is shown by $A=[A_{ij}]$, a $|V|\times|V|$ matrix where $A_{ij}=1$ if there is an edge between $V_i$ and $V_j$ in $G$. 


The structure of the rooted binary tree $T$ is very important in the encoding process. We will divide the organization of $T$ into different levels (also known as depth). 
Let $[\ell, i]$ denote the $i$th node of the $\ell$-th level of $T$. Note that the root node is $[0,1]$. Figure~\ref{GtoT.fig} shows our naming convention for all the other nodes. A node in $T$ will can contain multiple vertices from $G$. The cardinality of the node $[\ell, i]$ is shown with $|[\ell, i]|$. By default, $|[0, 1]| = |V|$.

\begin{defn}
A node $[\ell, i]$ is a \emph{left node} if it is the left child of a node in level $\ell -1$. In our convention, left nodes have odd index value $i$.
\end{defn}

\begin{defn}
A node $[\ell, i]$ is a \emph{right node} if it is the right child of a node in level $\ell -1$.  In our convention, right nodes have even index value $i$.
\end{defn}

\begin{defn}
For a level $\ell$, the \emph{first nonempty node} refers to node $[\ell, \alpha]$:
\begin{equation}\label{fne.eq}
[\ell, \alpha]=
\begin{cases}
[\ell, 1], \text{ if $|[\ell, 1]| > 0$}\\
[\ell, 2], \text{ otherwise}.
\end{cases}
\end{equation}
\end{defn}

Often times, we may wish to explicitly refer to the parent or children node of $[\ell, i]$. We use $[\ell, i].\node{parent}$ to refer to the parent node of $[\ell, i]$. We use $[\ell, i].\node{left}$ and $[\ell, i].\node{right}$ to refer to left and right child of node $[\ell, i]$. In the example tree shown in Figure~\ref{GtoT.fig}, node $[1,1] = [2, 1].\node{parent}$, node $[2,3] = [1,2].\node{left}$, and node $[2,4] = [1,2].\node{right}$.

Let $\mathcal{R}([\ell, i])$ denote the path from node $[\ell, i]$ to the root node. 
We find levels where $\mathcal{R}([\ell, i])$ \emph{and} $\mathcal{R}([\ell, \alpha])$ both contain left nodes and store those levels in $\mathcal{CI}([\ell, i])$. We also find levels where $\mathcal{R}([\ell, i])$ \emph{or} $\mathcal{R}([\ell, \alpha])$ contain left nodes and store those levels in $\mathcal{I}([\ell, i])$.

For example in Figure~\ref{GtoT.fig}, $\mathcal{R}([2,3]) = \{[2,3], [1,2], [0,1]\}$ and $\mathcal{R}([2,1]) = \{[2,1], [1,1], [0,1]\}$. We can see that $\mathcal{R}([2,3])$ and $\mathcal{R}([2,1])$ (node $[2,1]$ refer to $[2,\alpha]$) both have left nodes at level $\ell=2$ (i.e., nodes $[2,3]$ and $[2,1]$, respectively) and therefore, we get $\mathcal{CI}([2, 3])=\{2\}$. We can also verify that $\mathcal{I}([2,3])=\{1,2\}$ as node $[1,1]$ is another left node in $\mathcal{R}([2,1])$.
\begin{figure}[tbh]
	\begin{centering}
  \includegraphics[width= 1.5 in]{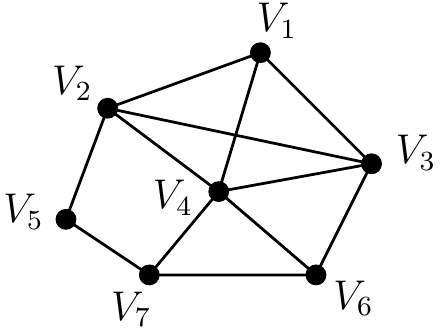}
  \caption{\label{Graph.fig}An example graph $G$ with $|V|=7$.}
 \end{centering}
\end{figure}

\begin{figure*}[tbh]
	\begin{centering}
  \includegraphics[width= 7 in]{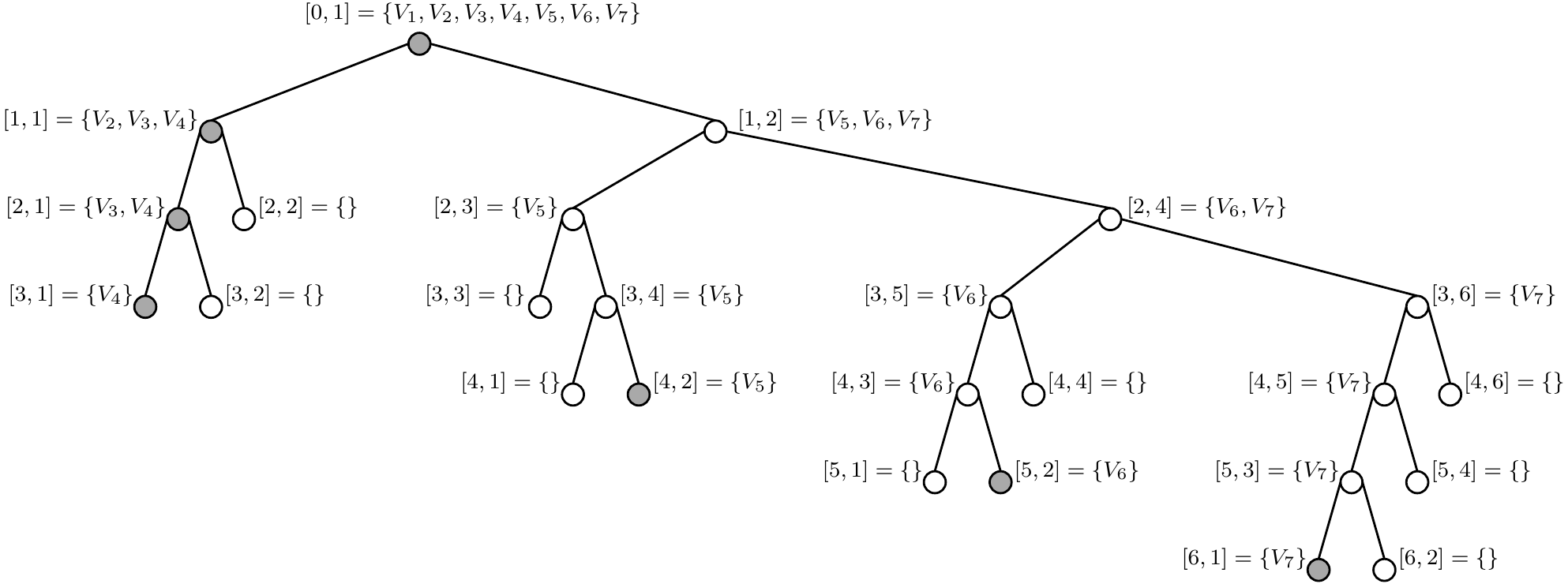}
  \caption{\label{GtoT.fig}Binary tree representation $T$ of the example graph $G$ in Figure~\ref{Graph.fig}. The root contains all vertices of $G$ and next levels are built by branching into neighbors and non-neighbors of vertex $V_i$ among each subset of vertices. The pair of $[\ell, i]$ next to each node shows the level that the node belongs to, $\ell$, and the position of the node in that level $i$ is determined by counting from the left to the right.}
 \end{centering}
\end{figure*}

\subsection{Transforming Graph $G$ into Binary Tree $T$}
The rooted binary tree $T$ is built by traversing the unlabeled graph $G(V,E)$. Each node in $T$ corresponds to a set of vertices from $G$. The root node $[0,1]$ corresponds to the set of all the vertices, $V$. 
To build levels $\ell= 1,\ldots, |V|-1$ of the tree we pick a vertex $V_{k(\ell)}$; we will specify
shortly how it is picked. Vertex $V_{k(\ell)}$ is removed
from the set of vertices. The remaining vertices belonging to each node at level $\ell-1$ is split into
two groups: the left node are all vertices connected to $V_{k(\ell)}$, and
the right node contain those not connected to $V_{k(\ell)}$. Empty nodes are not
split further. 
It is worth noting that the vertex $V_{k(\ell)}$  is picked randomly from the first non-empty node at
level $\ell-1$, i.e., node $[\ell-1, \alpha]$.
Figure~\ref{Graph.fig} and Figure~\ref{GtoT.fig} show an example of $G(V,E)$ and its corresponding rooted binary tree representation, $T$, respectively. Since the order of nodes
is irrelevant for the structure, here we have
assumed $V_{k(\ell)}=V_\ell$.

Algorithm~\ref{Alg.GtoT} gives the pseudocode for transforming graph $G$ into the rooted binary tree $T$. Note that hereafter, when we traverse a level in the rooted binary tree $T$, we start from the most left node and finish with the most right node at that level.

\begin{algorithm}
	\caption{Transform graph $G$ into rooted binary tree $T$}
	\label{Alg.GtoT}
    \begin{algorithmic}[1]
    \Function{GraphToTree}{$G$}
		\State {Create $root$ of $T$ with all vertices in $G$}
		\For {$\ell \gets 0$ to $|V|-2$}
		\State {Remove a vertex, $V_{\ell+1}$, randomly from node $[\ell, \alpha]$} 
		    \ForEach {node $[\ell,i]$ in the $\ell$th level of $T$}
		    \If {$[\ell,i] \not= NIL$}
		    \State{$[\ell,i].\node{left} \gets \text{neighbors of } V_{\ell+1}$ in $[\ell,i]$} 
		    \State{$[\ell,i].\node{right} \gets \text{non-neighbors of } V_{\ell+1}$ in $[\ell,i]$}
		    \EndIf
		    \EndFor
		\EndFor
		\State \Return $T$
	\EndFunction
	\end{algorithmic}
\end{algorithm}

\subsection{Encoding the Rooted Binary Tree $T$}

\begin{figure*}[tbh]
	\begin{centering}
  \includegraphics[width= 5.5 in]{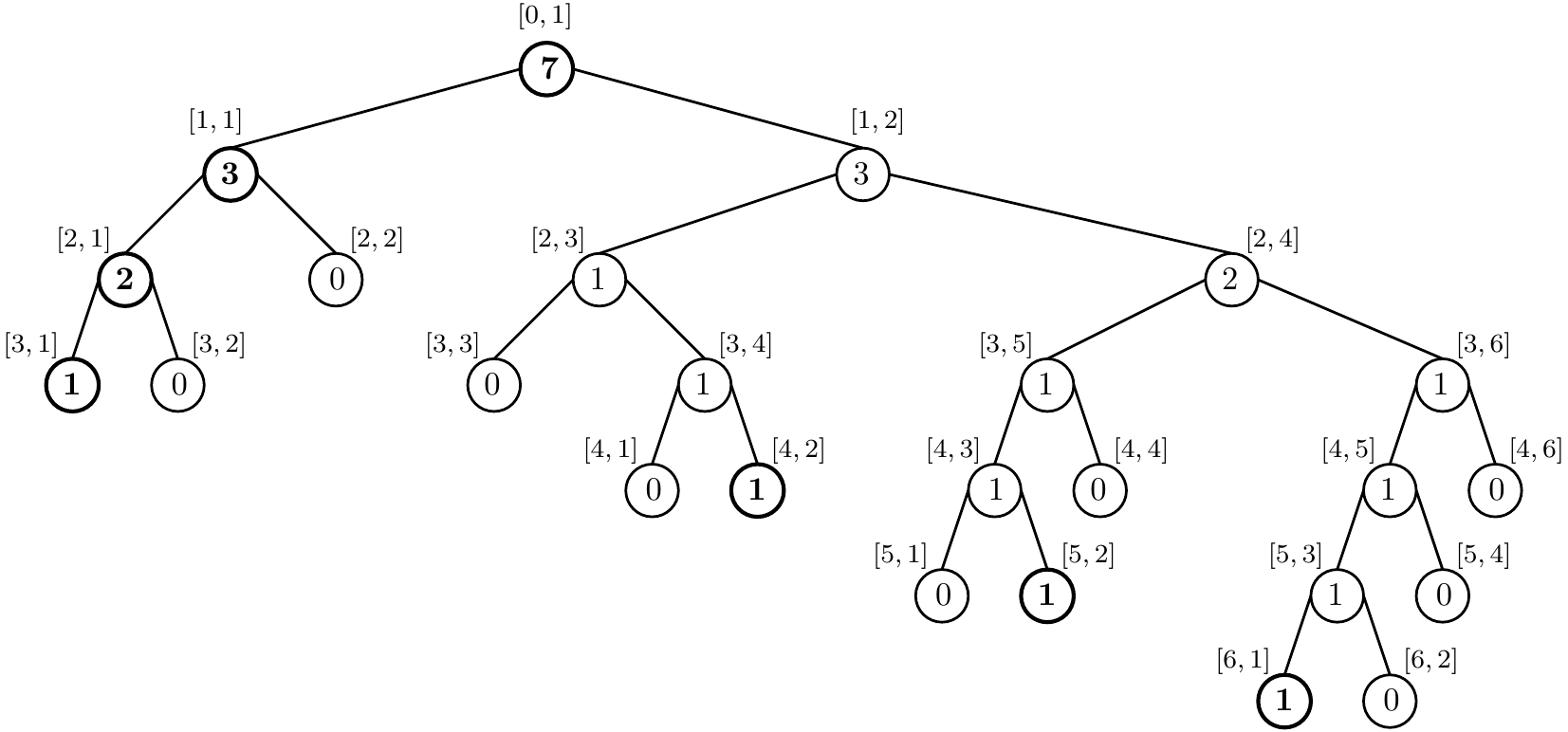}
  \caption{\label{Tree_tilde.fig}Transforming $T$ in Figure~\ref{GtoT.fig} into $\widetilde{T}$. Each node's value represents the cardinality of the the corresponding node in $T$.}
 \end{centering}
\end{figure*}

As we are only interested in coding the structure of $G$ (i.e., vertices up to an automorphism), we do not need to explicitly encode the set of vertices corresponding to the nodes in $T$. It is sufficient to encode only the cardinality of nodes; we call
the tree with cardinality as node values $\widetilde{T}$ (see Figure~\ref{Tree_tilde.fig}). When we refer to encoding a node $[\ell, i]$, we are referring to encoding the value of the node.

Furthermore, we \emph{only} need to encode the left nodes as the value of the siblings (i.e., right nodes) can be deduced given the value of the parent nodes. Consider a nonempty node $[\ell, i]$, we can see that 
\begin{equation}
|[\ell, i]| =  
\begin{cases}
|[\ell, i].\node{left}| + |[\ell, i].\node{right}|+1, \text{ if $[\ell,i]=[\ell, \alpha]$}\\
|[\ell, i].\node{left}| + |[\ell, i].\node{right}|, \text{ otherwise}.
\end{cases}
\end{equation}
The reason for the discrepancy is because of our convention of always removing a vertex to encode at from the first nonempty node $[\ell, \alpha]$ at each level.

It is easy to see that once the decoder has reconstructed the 
tree $\widetilde{T}$ and consequently $T$, one can reconstruct graph $\widetilde{G}$ isomorphic to the original graph $G$.
The procedure is very similar to the one for transforming a graph into a binary tree. 
We start with $|V|$ isolated vertices as determined by the cardinality of the root in $T$.
Then for each level $\ell=1,\ldots, |V|-1$, we connect the vertex $V_{\ell}$ to vertices in left nodes at level $\ell$ of $T$. The resulted graph $\widetilde{G}$ is isomorphic to the original graph $G$.

The remaining problem now is encoding the values in the tree $\widetilde{T}$.
The better the values can be predicted, the shorter the codelength. 
The encoder/decoder can predict the value either based on global
properties of the graph (e.g., degree distribution), or based on
local properties, which can depend only on the part of the 
tree $\widetilde{T}$ already encoded/decoded.
One such property is that $|[\ell, i]|$ can only take on integer values $0,1,2 ,\ldots, |[\ell, i].\node{parent}|-1, \text{ if $i = 1, 2$}$ or $0,1,2 ,\ldots, |[\ell, i].\node{parent}|, \text{ if $i > 2$}$. 

It is important to realize that the properties used for encoding
the tree $\widetilde{T}$ should be properties of the original graph,
since they presumably have statistical significance. As an example, if the original graph
is modeled as \ER\ \cite{bollobas2001random}, it is easy to see that the node values are
binomially distributed, $Binom(N, p)$, where
\begin{equation}\label{bin.eq}
N=
\begin{cases}
|[\ell, i].\node{parent}| -1, \text{ if $i = 1,2$}\\
|[\ell, i].\node{parent}|, \text{ if $i > 2$}.
\end{cases}
\end{equation}
The encoder uses the global property $p$, which
must be known by the decoder. In universal
coding, this can be transmitted to the
decoder initially. Section \ref{Calc.sec} outlines how to calculate
and encode global properties.
We call this approach IID coder, as it is based
on the i.i.d property of the \ER\ graph.
It is about equivalent to the approach introduced in \cite{ChoiSzpankowski12} for the compression of \ER\ graphs. The pseudocode for IID coder is shown in Algorithm~\ref{Alg.IID}.

\begin{algorithm}
	\caption{Encode $\widetilde{T}$ with IID coder}
	\label{Alg.IID}
    \begin{algorithmic}[1]
    \Function{EncodeIID}{$\widetilde{T}, p$}
        \State{Encode $|[0,1]|$ via a positive integer encoder}
		\For {$\ell \gets 1$ to $|V|-1$}
		    \ForEach{left node $[\ell, i]$}
		    \State{Encode $|[\ell,i]| \sim Binom(N,p)$ with $N$ from \eqref{bin.eq}} 
		    \EndFor
		\EndFor
	\EndFunction
	\end{algorithmic}
\end{algorithm}

In practice, \ER\ graphs are not good models for real-world graphs as edges are usually not independent \cite{albert2002statistical}. Therefore, 
we would like to use more advanced graph properties. We classify our coding methods broadly into two classes: 1) Node-by-Node Coder and 2) Level-by-Node Coder. In Node-by-Node Coder, we still use binomial distribution; however, the edge probability, $p$, will depend on local
properties (i.e., graph motifs) in $G(V,E)$. In Level-by-Node Coder, we use the degree distribution of $G(V,E)$ as a global property of graph to encode the values of all left nodes in level $\ell$ of $\widetilde{T}$ at the same time.

The value of a left node in $\widetilde{T}$ can equivalently
be seen as a count of \emph{edges}
of the original graph. Since only the \emph{number} of edges are encoded,
any local property used for encoding must be shared by all edges in
a left node. Equivalently, any property of the original
graph used for encoding must be convertible
to a property of the tree $\widetilde{T}$. In other words,
we convert the original graph $G(V,E)$ into the tree
$\widetilde{T}$, and any properties that we want to use for coding
must then become properties purely of $\widetilde{T}$.
We will describe this with more details for each class of coders.

\subsection{\label{EdgeProp.sec} Class 1: Node-by-Node Coder}
For Node-by-Node Coders, we will traverse $\widetilde{T}$ back to the root to determine the existence of certain motif structures in $G$. This will help us to better encode graphs that are not \ER\ as in these cases, edges are not independent of one another. 

\subsubsection{Coding of Triangles \label{Tri.sec}}
The first Node-by-Node Coder we consider is the triangle coder. The triangle coder results in shorter codelength for graph classes that induce more triangle structures such as scale-free graphs. A triangle is a cycle graph with three nodes, which is also a 3-clique. Statistics about triangles are often used to characterize graphs \cite{barabasi2016network}. 

First, we describe how to deduce the existence of triangles in $G$ from the structure of $\widetilde{T}$.
We know that the set of vertices corresponding to a left node $[\ell, i]$ are connected to the vertex $V_{\ell}$. We can also deduce if there are edges between the set of vertices corresponding to the left node $[\ell, i]$ and the vertex $V_{\ell-i}, i=1,\ldots, \ell-1$. For that, we look at the ancestor of node $[\ell, i]$ in level $\ell-i$. If it is a left node, then the set of vertices corresponding to the left node $[\ell, i]$ are connected to the vertex $V_{\ell-i}$. To have a triangle, there must be an edge between $V_\ell$ and $V_{\ell-i}$. It can be verified if the ancestor of $[\ell-1, \alpha]$ at level $\ell-i$ is a left node. 
If that is the case, we can deduce a triangle forms  between any of vertices corresponding to the left node $[\ell, i]$, $V_\ell$, and $V_{\ell-i}$. 
For example, consider node $[4,3] = \{V_6\}$ in Figure~\ref{GtoT.fig}. Since node $[4,3]$ is a left node in level $4$, there is an edge between $V_6$ and $V_4$. We can see that $V_6$ is also connected to $V_3$ because its ancestor in level $3$ (i.e., node $[3,5]$) is a left node. We can also verify that $V_4$ and $V_3$ are connected since node $[3,1]$ (i.e., node $[3, \alpha]$) is a left node at level $3$ and therefore, connected to $V_3$. Considering all these connections, we can deduce that $V_6, V_4, V_3$ form a triangle subgraph in $G$. We can use the formation of triangles to encode the nodal values in $\widetilde{T}$ as follows. 

Similar to IID coder, we use binomial distribution. However, the triangle coder chooses between two binomial distributions:
$Binom(N,\widecheck{p}_\triangle)$ or $Binom(N,p_\triangle)$, where $N$ is given by equation~\eqref{bin.eq}. We decide between these two distributions with the help of $\mathcal{CI}([\ell, i].\node{parent})$.
Consider a left node $[\ell, i]$, the coder will use $Binom(N,p_\triangle)$ to encode its value if 
$\mathcal{CI}([\ell, i].\node{parent})$ has at least one element. Note that $\mathcal{CI}([\ell, i].\node{parent})$ represents levels where $\mathcal{R}([\ell, i].\node{parent})$ and $\mathcal{R}([\ell-1, \alpha])$ both contain left nodes. It means that vertices corresponding to left node $[\ell, i]$ and $V_{\ell-1}$ are both connected to vertex $V_{\ell'}, \ell' \in \mathcal{CI}([\ell, i].\node{parent}$, and therefore, triangle forms among these vertices in $G$.
When $\mathcal{CI}([\ell, i].\node{parent})$ is empty, we encode $[\ell, i]$ with $Binom(N,\widecheck{p}_\triangle)$ as no triangle exists among these vertices in $G$. The psuedocode for the triangle coder is given in Algorithm~\ref{Alg.Tri}. 

\begin{algorithm}
	\caption{Encode $\widetilde{T}$ with triangles from Class 1}
	\label{Alg.Tri}
    \begin{algorithmic}[1]
    \Function{EncodeTriangles}{$\widetilde{T}, \{ \widecheck{p}_\triangle, p_\triangle \}$}
        \State{Encode $|[0,1]|$ via a positive integer encoder}
		\For{$\ell \gets 1$ to $|V|-1$}
		    \ForEach{left node $[\ell, i]$}
		    \If{ $|\mathcal{CI}([\ell,i].\node{parent})|>0$} 
		        \State{Encode $|[\ell,i]| \sim Binom(N,p_\triangle)$}
		    \Else 
		        \State{Encode $|[\ell,i]| \sim Binom(N,\widecheck{p}_\triangle)$} 
		    \EndIf
		    \EndFor
		\EndFor
	\EndFunction
	\end{algorithmic}
\end{algorithm}

\subsubsection{Coding with the number of common neighbors}

In a triangle subgraph, two connected vertices share a single common neighbor. But it may be that two vertices share multiple common neighbors as there is usually a correlation between having an edge between two vertices and the number of their common neighbors. This property is used for link prediction in complex networks \cite{yao2016link, li2018similarity}. Therefore, we can generalize the triangle encoder to encoding with $m$ common neighbors. Instead of using $\widecheck{p}_\triangle$ or $p_{\triangle}$  to parameterize the binomial distribution, we can use $p_{\triangle^{(m)}}$ where $m$ denote the number of common neighbors (note that $p_{\triangle^{(0)}}$ is the same as $\widecheck{p}_\triangle$ in coding with triangles). In other words, we find how many edges exist between any vertex corresponding to left node $[\ell, i]$ and vertex $V_\ell$ and utilize it as statistics that built the graph.
The number of common neighbors can range from $0$ to the maximum degree of a vertex in $G$, which we denote by $d_{max}$. 

The way we encode nodal values of $\widetilde{T}$ with this coder is as follows. 
For a left node $[\ell, i]$, we look at the cardinality of $\mathcal{CI}([\ell, i].\node{parent})$. It tells us how many vertices already encoded exist that any of vertices corresponding to $[\ell, i]$ and $V_\ell$ are both connected to them. This determines which parameter to pick for binomial distribution.

\subsubsection{Coding over 4-node motifs}
\begin{figure}[tbh]
	\begin{centering}
  \includegraphics[width=3.5 in]{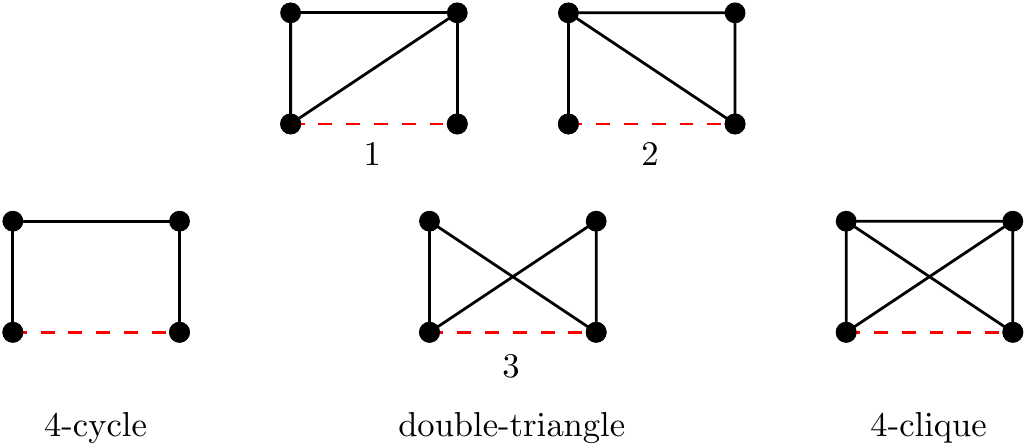}
  \caption{\label{4motifs.fig}Motifs of interest over four vertices with three realizations for double-triangle.}
 \end{centering}
\end{figure}

We can explore motifs of larger sizes in $G$ to develop new coders. However, computational cost can be a limiting factor. Here, we are interested to extend coding to motifs of size 4 vertices. Figure~\ref{4motifs.fig} illustrates motifs that we look for. Note that the priority starts with 4-clique. If 4-clique does not exist, then we look for double-triangle, and finally 4-cycle. 
Each of these motifs can be encoded with $p_{\boxtimes}$, $p_{\boxbslash}$, and $p_\square$, respectively. If none of them exist, then we encode left node $[\ell,i]$ using $\widecheck{p}_\square$.

As we mentioned earlier, we find the appropriate parameter to encode nodal values of $\widetilde{T}$ by looking at the structure of encoded/decoded part of tree $\widetilde{T}$ up to that point. Depending on the cardinality of $\mathcal{CI}([\ell, i].\node{parent})$, different cases are possible:

\begin{itemize}
    \item $|\mathcal{CI}([\ell, i].\node{parent})| \geq 2:$ For any pair $\ell_1$ and $\ell_2$ in $\mathcal{CI}([\ell, i].\node{parent})$ where $\ell_1 < \ell_2$, we check to see if we can find a case in which the first nonempty node in level $\ell_2$ (i.e., $[\ell_2, \alpha]$) lies in the left subtree of the tree rooted at each node in level $\ell_1$. If such a case exists, we use $Binom(N, p_\boxtimes)$ to encode the value of $[\ell, i]$. Otherwise, the configuration~3 of double-triangle in Figure~\ref{4motifs.fig} occurred; thus, we use $Binom(N, p_\boxbslash)$.
    
    \item $|\mathcal{CI}([\ell, i].\node{parent}| = 1:$ Two cases are possible
    \begin{itemize}
        \item For the first nonempty node in level $\ell_1, \ell_1 \in \mathcal{CI}([\ell, i].\node{parent})$, and node $[\ell-1, \alpha]$, we check to see if $\mathcal{R}([\ell_1, \alpha])$ and $\mathcal{R}([\ell-1, \alpha])$ contain a left node at the same level. If such a node exists, the configuration~1 of double-triangle in Figure~\ref{4motifs.fig} occurred; thus, we use $Binom(N, p_\boxbslash)$.
        \item For node $[\ell_1, \alpha]$ and $[\ell, i].\node{parent}$, we check to see if $\mathcal{R}([\ell_1, \alpha])$ and $\mathcal{R}([\ell, i].\node{parent})$ contain a left node at the same level. If such a node exists, the configuration~2 of double-triangle in Figure~\ref{4motifs.fig} occurred; thus, we use $Binom(N, p_\boxbslash)$.
    \end{itemize}
    
    \item $|\mathcal{CI}([\ell, i].\node{parent})| = 0:$ First, we need to find $\mathcal{I}([\ell, i].\node{parent})$ which contains all the levels that $\mathcal{R}([\ell,i].\node{parent})$ or $\mathcal{R}([\ell-1,\alpha])$ have left nodes. For $|\mathcal{I}([\ell, i].\node{parent})| > 1$ and any pair $\ell_1$ and $\ell_2$ in $\mathcal{I}([\ell, i].\node{parent})$ where $\ell_1 < \ell_2$, we look for a case in which the first nonempty node in level $\ell_2$ (i.e., $[\ell_2, \alpha]$) lies in the left subtree of the tree rooted at any node in level $\ell_1$. If such a case exists, we use $Binom(N, p_{\square})$ to encode the value of $[\ell,i]$.
    
    \item If none of the abovementioned cases occurred, we use $Binom(N, \widecheck{p}_\square)$ to encode the value of $[\ell,i]$.
    
\end{itemize}

Due to similarity, we skip the pseudocode for this coder to avoid repetition.

\subsection{\label{EdgeNodeProp.sec}Class 2: Level-by-Node Coder}
One important graph statistics is the degree distribution $P(k)$. Class 2 coders utilize the degree distribution in addition to graph motifs presented in Class 1 coders to efficiently encode $\widetilde{T}$. 

For a given level, $\ell$, in $\widetilde{T}$, Class 1 coders encode the value of each left node (and deduces the value of the right node) independently of one another. Class 2 coders on the other hand, utilize the degree distribution to encode the values of left nodes at the same level altogether.

The number of left nodes encountered in $\mathcal{R}([\ell-1, \alpha])$, $\widecheck{k}$, is equal to the number of edges, $V_{\ell} \leftrightarrow  V_{\ell - i}, i =1,\ldots, \ell-1$, in $G$. We know that the degree of vertex $V_{\ell}$ is lower bounded by $\widecheck{k}$. The sum of the values of left nodes in level $\ell$ is $k - \widecheck{k}$, where $k$ is the degree of vertex $V_{\ell}$.
We can use this relationship to encode all the left nodes in level $\ell$ of $T$ at the same time.

Encoding with Class 2 entails two steps. First, we need to send the degree of $V_{\ell}$. Then, we send the conditional probability of seeing specific values for left nodes at level $\ell$ conditioned on their summation to be $k - \widecheck{k}$. 

For the first step, we use the degree distribution $P(k)$ and the fact that $k \geq \widecheck{k}$. Thus, we encode the degree of $V_{\ell}$ with
\begin{equation}
P(k|k\geq \widecheck{k})=\frac{P(k)}{\sum\limits_{j \geq \widecheck{k}}P(j)}.\label{Pk2.eq}
\end{equation}
One should notice that at the time of decoding a given level, the decoder knows $\widecheck{k}$ based on the tree up that point and thus, it can calculate (\ref{Pk2.eq}).
Now, the decoder can compute the summation of the left nodes' values at level $\ell$ with
\begin{equation}\label{eq:levelsum}
\sum_{i \text{ is odd}} |[\ell, i]| = k - \widecheck{k}.
\end{equation}
Let $[\ell, N_\ell]$ denote last left node in level $\ell$. To encode left nodes' values in level $\ell$, we need to compute the joint probability of observing specific values for left nodes at level $\ell$ (i.e., left nodes take values $|[\ell,1]|, \ldots, |[\ell, N_\ell]|$ respectively) conditioning on their summation being equal to \eqref{eq:levelsum}. 
\begin{align}\label{eq:class2}
    P\big( |[\ell,1]|,\ldots, &|[\ell, N_\ell]|\ \big|  {\textstyle \sum\limits_{i \text{ is odd}}} |[\ell, i]| = k - \widecheck{k} \big) \nonumber\\
    & = \frac{P(|[\ell,1]|,\ldots, |[\ell, N_\ell]|)}{P\big(\sum\limits_{i \text{ is odd}} |[\ell, i]| = k - \widecheck{k}\big)}  \nonumber\\
    &= \ddfrac{\prod\limits_{i=1}^{N_{\ell}}P(|[\ell, i]|)}{P\big(\textstyle\sum\limits_{i \text{ is odd}} |[\ell, i]| = k - \widecheck{k}\big)},
\end{align}
where the numerator is the joint probability of observing $|[\ell,1]|,\ldots, |[\ell, N_\ell]|$ as the values of left nodes and the denominator is the probability corresponding to all configurations with the same summation for left nodes' value, $k - \widecheck{k}$.
By independent assumption on the values of left nodes, the joint probability distribution $P(|[\ell,1]|,\ldots, |[\ell, N_\ell]|)$ will reduce to the product of individual probabilities.

Assuming the independence assumption, two cases are possible:

\begin{enumerate}
    \item \emph{Identically distributed:}
    This case happens when we want to compute \eqref{eq:class2} using IID coder. Then, the probability of success is the same for all probabilities and the problem reduces to a counting problem. We can encode the values of left nodes using the probability distribution
\begin{align} \label{eq:deg}
    P\big( |[\ell,1]|,\ldots, &|[\ell, N_\ell]|\ \big| {\textstyle \sum\limits_{i \text{ is odd}}} |[\ell, i]| = k - \widecheck{k} \big) \nonumber\\
    &= \ddfrac{\prod_{i \text{ is odd}} { |[\ell, i]|+|[\ell, i+1]| \choose |[\ell,i]|} }{{\sum\limits_{i} |[\ell, i]| \choose \sum\limits_{i \text{ is odd}} |[\ell, i]|}} \nonumber\\
    &= \ddfrac{\prod_{i \text{ is odd}} { |[\ell, i]|+|[\ell, i+1]| \choose |[\ell,i]|}}{{\sum\limits_{i} |[\ell, i]| \choose k -\widecheck{k}}}.
\end{align}
\item \emph{Not identically distributed:}
If we use any other coders except IID coder, then different left nodes in the same level \emph{do not} have the same parameter for encoding. For example as we showed with the triangle coder from Class 1, node $[2,1]$ in Figure~\ref{Tree_tilde.fig} would be encoded with $p_\triangle$ while node $[2,3]$ would be encoded with $\widecheck{p}_\triangle$. 
To take this into account, we need to encode the values of left nodes using the following probability distribution
\begin{align}\label{eq:pbd}
    P\big( &|[\ell,1]|,\ldots, |[\ell, N_\ell]|\ \big| {\textstyle \sum\limits_{i \text{ is odd}}} |[\ell, i]| = k - \widecheck{k} \big) \nonumber\\
    &= \ddfrac{\prod_{i \text{ is odd}} { |[\ell, i]|+|[\ell, i+1]| \choose |[\ell,i]|} \theta^{|[\ell,i]|} (1 - \theta)^{|[\ell, i+1]|} } {P\big({\textstyle  \sum\limits_{i \text{ is odd}}} |[\ell, i]| = k - \widecheck{k}\big)},
\end{align}
where $\theta$ is the relevant parameter for each left node in the tree $\widetilde{T}$ (e.g., $p_\triangle$ or $\widecheck{p}_\triangle$ when using triangles for coding). 

The denominator in \eqref{eq:pbd} is a generalized version of binomial distribution called Poisson binomial distribution \cite{wang1993number}. It is defined as the sum of independent binomial distributions that are not necessarily identically distributed.
Due to the involvement of different probabilities, calculation of denominator in \eqref{eq:pbd} can be cumbersome. To resolve this, some methods were proposed in the literature.
Recursive methods were developed in \cite{chen1994weighted,Barlow1984computing} that can compute the denominator in \eqref{eq:pbd} in $O\big((k - \widecheck{k}) \textstyle \sum_{i} |[\ell, i]| \big)$ time.
We pay this extra computational cost to have a more efficient coder since the entropy of Poisson binomial distribution is bounded above by the entropy of binomial distribution with the same mean \cite{harremoes2001binomial}. The pseudocode to encode with Class 2 is given in Algorithm~\ref{Alg.Class2}.
\end{enumerate}

\begin{algorithm}
	\caption{Encode $\widetilde{T}$ with Class 2}
	\label{Alg.Class2}
    \begin{algorithmic}[1]
    \Function{EncodeClass2}{$\widetilde{T}$}
        \State{Encode $|[0,1]|$ via a positive integer encoder}
		\For{$\ell \gets 1$ to $|V|-1$}
		    \State{$\widecheck{k} \gets$ Number of left nodes in $\mathcal{R}([\ell-1, \alpha])$} 
		    \State{$\widehat{k} \gets$ Summation of left nodes' values in level $\ell$}
		    \State{$k \gets \widecheck{k}+ \widehat{k}$}
		    \State{Encode $k$ with \eqref{Pk2.eq}}
		    \State{$P \gets$ Compute \eqref{eq:class2}}
		    \State{Encode $P$}
		\EndFor
	\EndFunction
	\end{algorithmic}
\end{algorithm}

\subsection{\label{Calc.sec}Calculation and Encoding of Statistics}
We consider encoding in two scenarios: learned coding, and universal coding. In learned coding, we are given a set of training graphs $\{G_1,\ldots, G_N\}$ of a particular class and have to learn local and global statistics; these statistics, then, are shared to both encoder and decoder. In universal coding, there is no training set and the encoder encodes a single graph. It also has to communicate to the decoder what is the statistics.
Below we describe calculation and communication of statistics for each of these scenarios.

\subsubsection{Learned Coding}
For learned coding, we need to learn statistics from a set of training graphs. To do that, each statistic is calculated by taking an average over the same statistic in the training set. The edge probability $p$ in coding with IID coder can be estimated by the average degree. Other edge statistics are more tricky and should reflect the procedure used for encoding the graph. It means that we cannot simply count the number of triangles in a graph to compute $p_\triangle$. The reason is when we want to encode a graph with triangles, we look for the formation of a triangle in a specific way forced by the coding algorithm as described earlier.
Thus, we need to transform each  graph $G_i$ from the training set into its rooted binary tree representation $T_i$ and find statistics with coding algorithm. To estimate $p_\triangle$ and $\widecheck{p}_\triangle$, we traverse each $T_i$ the same way we did for coding and divide the tree's nodes into those coded with $p_{\triangle}$ and those coded with $\widecheck{p}_\triangle$. To estimate each of $p_{\triangle}$ and $\widecheck{p}_\triangle$, we compute the ratio of the summation of left nodes' values to the summation of left and right nodes' values in that group. The average over all $T_i$ gives the estimation for $p_\triangle$ and $\widecheck{p}_\triangle$. The same recipe is used to estimate edge statistics for common neighbors and 4-node motifs.

The estimation of degree distribution in Class 2 is straightforward. It can be estimated through the histogram. We compute the degree distribution for each graph in the training set and the final degree distribution is estimated by taking average over them.

\subsubsection{Universal Coding}
For encoding average degree in IID coder, we can send the number of edges $|E|$ in $G$. The number of bits required to encode the number of edges is about $\log\frac{n(n-1)}{2}\approx2\log n$ bits. Once the decoder knows the number of edges, it can compute the parameter of  IID coder, $p$, with
\begin{equation*}
    p= \frac{2|E|}{n(n-1)}.
\end{equation*}

For other local statistics, we use sequential estimation best outlined in \cite[Section 13.2]{CoverBook}. For example in coding with triangles, we use sequential estimation of $p_{\triangle}$ and $\widecheck{p}_\triangle$, specifically the KT estimator \cite{KrichevskyTrofimov81,WillemsAl95}, which is
\begin{equation*}
    \hat{p}=\frac{n_{1}+\frac{1}{2}}{n_{1}+n_{0}+1},
\end{equation*}
where $n_{1},n_{0}$ are the summation of left and right nodes' values previously coded in the rooted binary tree, respectively. 
One should note that the procedure for updating the probabilities $p_{\triangle}$ and $\widecheck{p}_\triangle$ is different for each class.
For Class 1 coders, the probabilities will be updated after coding each node of the rooted binary tree. However, for Class 2 coders, the probabilities will be updated after coding each level. The reason is that all nodes at the same level will be encoded together.
To estimate statistics for common neighbors and 4-node motifs, we utilize similar sequential approach used for the estimation of $p_\triangle$ and $\widecheck{p}_\triangle$.

For the degree distribution, we calculate the degree histogram for the whole graph, and use this for coding. The degree of a node is between 0 and $n-1$. We can therefore think of the degree histogram as putting each of the $n$ (unlabeled) nodes into one of $n$ buckets, and encoding this can be done by encoding the counts in the buckets. The number of possible configurations is a standard problem in combinatorics: ${2n-1 \choose n}$, which can be transmitted with
\begin{align*}
    \log {2n-1 \choose n} & = nH\left(\frac{n}{2n-1}\right)+\frac{1}{2}\log\frac{2n-1}{n^{2}}+c \\
    & \approx n-\frac{1}{2}\log n\ \text{bits} \ (|c|\leq2).
\end{align*}

Hereinafter, in our experiments, we use the above mentioned approach to calculate and communicate local and global statistics of graphs.

\subsection{\label{GraphCom_Exp} Experiments}
We consider two cases for experiments: 1) using learned coding to encode a graph of the same class of training set, 2) using universal coding to encode a single graph. In the former case, we generate synthetic data for different classes of graphs, whereas in the later case we measure the performance on real-world graphs. We evaluate the performance of our coding methods versus IID coder and also compare the performance of Class 1 and Class 2 against each other.

\begin{figure*}[tbh]
\begin{centering}
\includegraphics[width=7in]{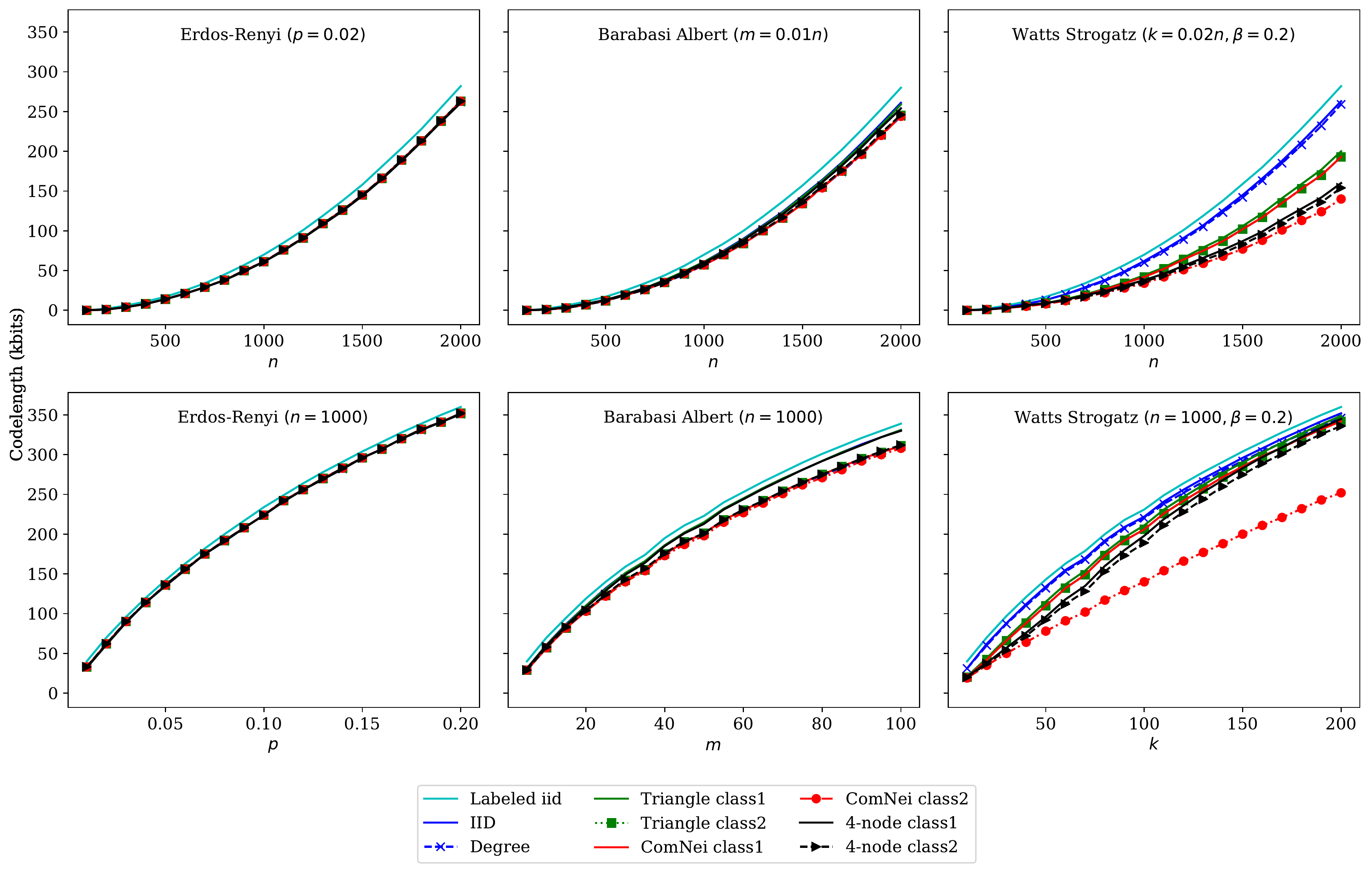} 
\par\end{centering}
\caption{\label{GraphComp.fig} Comparison of codelength associated with each coder for \ER, \BA, and Watts Strogatz graphs. Left plots are for the graphs of variable size and the average degree of $\langle k \rangle = \frac{n}{50}$. Right plots are for graphs of the same size $n=1000$ and $\langle k \rangle$ ranging from $\frac{n}{100}$ to $\frac{n}{5}$. In \BA\ graph, $m$ denote the number of new edges added by each node. In Watts Strogatz, $k$ is the number of nearest neighbors in the model and $\beta$ refer to rewiring probability.}
\end{figure*}

For learned coding, we first need to learn statistics for coders. We considered different classes of graphs in our experiments. In all cases, learning was done on 50 graphs. Then, we used those statistics from the training to encode a test graph of the same type. 
The results are shown in Figure~\ref{GraphComp.fig}.
As expected, for \ER\ graph, 
IID coder is efficient and all other coders do not offer an improvement. However, for \BA\ and Watts Strogatz graphs, our proposed coders outperform IID coder by a significant margin. One can observe that coders in Class 2 have shorter codelength than their counterparts in Class 1. For \BA\ graph, all encoders in Class 2 almost have the same performance and therefore, choosing one or another does not matter. However, for Watts Strogatz graph, coding with common neighbors through Class 2 is the most efficient. 

For compression of real-world graphs, we use universal coding since there is no training set. Note that the statistics associated with each coder is required to be communicated to the decoder as described in Section~\ref{Calc.sec}.
We measure the compression performance of Class 1 and Class 2 coders against each other and also compute the codelength associated with IID coder and arithmetic coding (represented by labeled iid) for comparison. The following undirected graphs, publicly available to download \footnote{\url{http://networkrepository.com/}}\textsuperscript{,}\footnote{\url{https://snap.stanford.edu}}, are considered:
\begin{itemize}
    \item Genetic interaction: a gene interaction network \cite{nr}.
    \item Economic Network: the economic network of Victoria, Australia in 1880 \cite{nr}.
    \item Transportation network: A simple graph of Minnesota road network \cite{nr}.
    \item Collaboration network: extracted from \emph{arXiv} papers and covers collaborations among authors in General Relativity and Quantum Cosmology category \cite{leskovec2007graph}.
    \item Facebook-politician: represents blue verified Facebook page networks of politicians \cite{rozemberczki2019gemsec}.
    \item Internet (AS level): An undirected graph representing AS peering information inferred from  University of Oregon Route Views Project on January 1, 2000 \cite{leskovec2005graphs}.
\end{itemize}

As we can see from Table~\ref{RealGraphs.tab}, coders in Class 2 give a better performance compared to their counterparts in Class 1. This improvement comes at the cost of computing \eqref{eq:pbd}. Moreover, shorter codelength is achievable by using Class 1 over IID coder. Another observation is that coding with common neighbors by means of Class 2 outperforms other coders in most cases. However, there are some cases that coding over 4-node motifs and triangles offer improvement over common neighbors.
There is no single coder that is best for all graphs.
This is no different than the situation for encoding
sequences. If one wants the most efficient coder
for a particular graph, one can encode with
multiple algorithms, and choose the shortest, indicating
by a header which is used or use soft combining
\cite{VolfWillems98}. In the end it is a 
tradeoff between compression and complexity.

\begin{table*}[tbh]
\vspace{0.5cm}
\begin{center}
\setlength{\tabcolsep}{4pt}
\renewcommand{\arraystretch}{1.4}
\begin{tabular}{ |c|c|c|c|c|c|c|c|c|c|c|c| } 
 \hline
 \multirow{2}{*}{Name} & \multirow{2}{*}{$|V|$} & \multirow{2}{*}{$|E|$} & \multirow{2}{*}{Labeled iid} & \multirow{2}{*}{IID coder} &\multicolumn{3}{c|}{Class 1} & \multicolumn{4}{c|}{Class 2}  \\
 \cline{6-12}
 & & &  & & Triangle & ComNei & 4-node & Degree & Triangle & ComNei & 4-node \\
 \hline
 
 Genetic interaction & 4227 & 39484 & 365599 & 317379 & 273444 & 235011 & 249467 & 280375 & 235787 & \textbf{206485} & 220191\\
 Economic network & 1258  & 7513 & 61224 & 47592 & 49790 & 49597 &  46407 & 45500 & 44159 & 43996 & \textbf{40730} \\
 Transportation network & 2642 & 3303 & 37937 & 35424 & 14619 & 14620 & 14596 & 12712 & \textbf{12530} & 12531 & 12540 \\
 Collaboration network & 5242 & 14496 & 164224 & 119038 & 60375 & 52041 & 66203 & 106215 & 53647 & \textbf{47320} & 60182 \\
 Facebook-politician & 5908 & 41729 & 423451 & 340070 & 222500 & 199514 & 206720 & 327203 & 195739 & \textbf{172386} & 179836 \\
 Internet (AS level) & 3570 & 7750 & 86206 & 47685 & 48266 & 46665 & 47222 & 31107 & 30469 & \textbf{29708} & 30897 \\
 
 \hline
\end{tabular}
\end{center}
\caption{\label{RealGraphs.tab}Compression length (bits) for some real-world graphs. Labelled iid encodes the graph with its labels. Degree denote IID coder when we use it along degree distribution in Class 2.
The best value for each case is boldfaced.}
\end{table*}
\section{Model Selection in Gaussian Graphical Models \label{GLASSO.sec}}
In this section, we introduce an application of graph coding for model selection in Gaussian graphical models. The goal is to find a graph that describes dependencies among multivariate Gaussian variables accurately \cite{friedman2008sparse}. Multivariate Gaussian distributions are widely used in modeling real-world problems where the relationship among approximately normally distributed variables is of interest. They have been used to model different datasets such as stock returns \cite{kon1984models, golosnoy2012conditional}, protein-protein interactions \cite{baldassi2014fast}, and brain functional activities \cite{varoquaux2010brain}.

Let $X = [X_1, \ldots, X_p]^T$ be a $p-$dimensional random vector with multivariate Gaussian distribution $\mathcal{N}(0,\Sigma)$. The inverse of covariance matrix, $\Sigma^{-1} = \Omega$, is known as the precision matrix. 
If the $(i,j)$ entry of the precision matrix (i.e., $\Omega_{ij}$) is $0$, then $X_i$ and $X_j$ are conditionally independent given all the other variables. Therefore, we can visualize the conditional independence relationships between any pair of variables as an unweighted, undirected graph, $G$ whose adjacency matrix, $A$ is such that 
\begin{equation}
\label{eq:Adj}
A_{ij}=
\begin{cases}
1 \hspace{0.5cm} &\text{if} \hspace{0.3cm} \Omega_{ij}\neq 0, i \neq j \\
0 \hspace{0.5cm} &\text{otherwise}.
\end{cases}
\end{equation}

The graph $G$ is known as the conditional independence graph. Its structure is determined by the sparsity pattern of $\Omega$.

Often, we are interested in the problem of estimating $G$ from $N$ i.i.d observations. To do that, we need to estimate $\Omega$ first.
The problem is especially difficult when $N < p$. In this case, the maximum likelihood estimate does not exist. In most approaches, the estimator of $\Omega$ contains a regularization term $\lambda$ to 1) prevent overfitting, 2) control the sparsity of the estimated solution. A popular sparse estimator of the precision matrix is found by maximizing the $L_1-$penalized log-likelihood \cite{friedman2008sparse, yuan2007model, banerjee2008model}
\begin{equation}\label{eq.glasso}
   \underset{\Omega \succ 0}{\text{max}} \hspace{0.2cm}  \mbox{logdet} \hspace{0.1cm} \Omega -\mbox{tr}(S\Omega) - \lambda \|\Omega\|_1,
\end{equation}
where $S$ is the sample covariance matrix, and $\lambda$ is the regularization parameter that controls sparsity. The performance of the estimator in \eqref{eq.glasso} is highly influenced by the selection of $\lambda$. Depending on the value of $\lambda$, the conditional independence graph can range from a dense graph for small values of $\lambda$ to a graph with zero edges when $\lambda$ takes large values. To find the best value of the regularization parameter $\lambda$, model selection techniques are being used.

We use the MDL principle and select $\lambda$ that minimizes the sum of the description length of the conditional independence graph structure and data when encoded under that graph structure
\begin{equation}
    \arg \min_{\lambda} L(G_{\lambda}) + L(D|G_{\lambda}),
    \label{eq:MDL}
\end{equation}
where $L(G_{\lambda})$ is the description length of the conditional independence graph $G_{\lambda}$, and $L(D|G_{\lambda})$ is the description length of data $D$ when encoded with $G_{\lambda}$. 

To compute the description length of the conditional independence graph, $L(G_{\lambda})$, we first need to find the underlying conditional independence graph $G_{\lambda}$. By applying an estimator to the sequence of $N$ i.i.d observations $\{x_1, \ldots, x_N\}$ from $p-$variate Gaussian distribution, we obtain an estimated precision matrix $\widehat{\Omega}$ for each realization of $\lambda$. We have dropped the explicit 
dependence on $\lambda$ from $\widehat{\Omega}$ and other $\lambda-$related precision and covariance matrices for the sake of simplicity of notation. In this paper, we use graphical lasso \cite{friedman2008sparse}, which is one of the most commonly used solvers of $L_1-$penalized log-likelihood in Gaussian graphical models. It is worth mentioning that our approach can be easily applied to any other estimator. The conditional independence graph $G_{\lambda}$ is obtained from the estimated precision matrix $\widehat{\Omega}$. We then pick a graph coder described in Section~\ref{Coding.sec} to compute the description length of $G_{\lambda}$.

To compute the description length of the data, $L(D|G_{\lambda})$, we face two challenges. The first challenge is to deal with real-valued data $\{x_1, \ldots, x_N\}$ where lossless source coding is not generalized directly. Second, we have to encode the data based on the underlying conditional independence graph $G_{\lambda}$. To encode real-valued data, we assume a fixed-point representation with a (large) finite number, $r$, bits after the period, and an unlimited number of bits before the period \cite{Rissanen83}. Therefore, the number of bits required to represent data $x$ according to the probability distribution function (pdf) $f(x)$ is given by
\begin{align}
L(x) &= - \log \int _{x}^{x+2^{-r}} f(t) dt \approx - \log \left(f(x)2^{-r} \right) \nonumber \\
 &= - \log f(x) + r.
\end{align}
Since we are only interested in relative codelengths between different models, the dependency on $r$ cancels out.

The second challenge is how to encode the data with respect to $G_{\lambda}$. As mentioned, the data is generated by multivariate Gaussian distribution, which can be characterized by covariance matrix $\Sigma$. Since we do not have access to the true covariance matrix $\Sigma$, we can find an estimate $\widecheck{\Sigma}$ based on the sample covariance matrix $S$. It should be noted that the structure of $\widecheck{\Sigma}^{-1}$ should match with the structure of $G_{\lambda}$ obtained from the previous step
\begin{equation} \label{demp.eq}
\begin{cases}
\widecheck{\Sigma}_{ij} = S_{ij} &\text{if} \hspace{0.3cm} i=j \ \text{or} \ A_{ij} \neq 0, \\
\widecheck{\Sigma}_{ij}^{-1} = 0 \hspace{0.5cm} &\text{otherwise}.
\end{cases}
\end{equation}
This problem is known as \emph{matrix completion} problem \cite{grone1984positive}. Dempster in \cite{dempster1972covariance}, proved the existence and uniqueness of maximum likelihood solution for $\widecheck{\Sigma}$ when the sample covariance matrix $S$ is positive definite. 
He presented a coordinate-descent algorithm to find the solution iteratively.
It should be noted that the precision matrix obtained from the graphical lasso solution $\hat{\Omega}$ is not necessarily equal to the inverse of estimated covariance matrix $\widecheck{\Sigma}$.

Once we estimate $\widecheck{\Sigma}$, we use predictive MDL \cite{Rissanen86} to compute $L(D | G_{\lambda})$

\begin{equation}
    L(D | G_{\lambda}) = -\sum_{i=0}^{N-1} \log f\left(x_{i+1}|\hat{\theta}(x_1,\ldots, x_i)\right), \label{eq:predMDL}
\end{equation}

where $f(\cdot|\cdot)$ is the conditional pdf and $\hat{\theta}(x_1,\ldots, x_i)$ denote the maximum likelihood estimate of parameters, which in this case is the estimated covariance $\widecheck{\Sigma}$ obtained under $G_{\lambda}$. The codelength in \eqref{eq:predMDL} is a valid codelength since it is sequentially decodable. Note that it does not work for the first few samples, as there is no estimate for $\widecheck{\Sigma}$. Instead, we encode the first few samples with a default distribution, which is the same among different realizations of $\lambda$. 

Finally, the best conditional independence graph structure $G_{\lambda^*}$ is obtained by minimizing $L(G_{\lambda}) + L(D|G_{\lambda})$ over $\lambda$. In the following, we present an algorithm to find the best graph model of data $G_{\lambda^*}$ associated with $\lambda^*$.

\begin{algorithm}
	\caption{Find the best graph model of data $G_{\lambda^*}$ via graph coding}
	\label{Alg}
	\hspace*{\algorithmicindent} \textbf{Input: }Samples $\{x_1, \ldots, x_N\} \sim \mathcal{N}(0,\,\Sigma)$ and regularization parameters $\{\lambda_1,\ldots, \lambda_K\}$. \\
    \hspace*{\algorithmicindent} \textbf{Output:} Best graph model of data $G_{\lambda^*}$.
    
    \begin{algorithmic}[1]
		\For {Each realization of $\lambda \in \{\lambda_1,\ldots, \lambda_K\}$.}
		\State {Apply graphical model estimator to find $\hat{\Omega}$ and build $G_{\lambda}$ based on \eqref{eq:Adj}.}
		\State {Use any graph coder in Section~\ref{Coding.sec} to compute $L(G_{\lambda})$.} 
		\State {Compute $L(D|G_{\lambda})$ via \eqref{eq:predMDL} where $\hat{\theta}(x_1,\ldots, x_i)$ denote $\widecheck{\Sigma}$ obtained from \eqref{demp.eq}.}
		\State {Add up codelengths resulted from step~(3) and step~(4).}
		\EndFor
		\State \Return {$G_{\lambda^*}$ associated with the shortest total codelength in step~(5).}
	\end{algorithmic}
\end{algorithm}

\subsection{Experiments\label{sec:ModelExp}}
We tested our approach on both synthetic and real-world data. For synthetic data, we consider different conditional independence graph structures and the goal is to recover true conditional independence graph. For real-world data, we applied our approach to ECG data of a group of healthy subjects and a group of subjects with Kawasaki disease. 
We are interested to determine if there is any difference between the conditional independence graph of healthy group versus Kawasaki group.

We now provide simulation results on synthetic data generated from zero-mean multivariate Gaussian distribution with known precision matrix, $\Omega$. We will compare the performance of graph coding methods against other methods in the recovery of conditional independence graph $G$. We use F1-score as a widely used metric in the literature. F1-score is the harmonic mean of precision and recall where the precision is the ratio of the number of correctly estimated edges to the total number of edges in the estimated conditional independence graph, and the recall is the ratio of the number of correctly estimated edges to the total number of edges in the true conditional independence graph \cite{liu2010stability}.
We consider two cases: 1) the number of observations is larger than the number of variables, $N>p$, with $N/p = 2$ and 2) the number of observations is smaller than the number of variables, $N<p$, with $N/p = 0.5$. The experiments were repeated for $p = 100, 200$.
We applied the graphical lasso estimator described in \cite{friedman2008sparse} for $\lambda \in [0.01, 1]$ when $N>p$ and $\lambda \in [0.1, 1]$ when $N<p$ to cover a wide range of structures from dense graphs to totally isolated nodes.

\begin{table*}[!tbh]
\begin{center}
\setlength{\tabcolsep}{4pt}
\renewcommand{\arraystretch}{1.4}
\begin{tabular}{ |c|c|c|c|c|c|c|c|c|c|c|c|c|c|c| } 
 \hline
 \multirow{2}{*}{Type} & \multirow{2}{*}{$p$} & \multirow{2}{*}{$N$} & \multirow{2}{*}{Optimum} & \multicolumn{3}{c|}{Benchmark methods} & \multirow{2}{*}{IID} & \multicolumn{3}{c|}{Class 1} & \multicolumn{4}{c|}{Class 2}  \\
 \cline{5-7} \cline{9-15}
 &  & & & CV & BIC & EBIC &  & Triangle & ComNei & 4-node & Degree & Triangle & ComNei & 4-node \\
 \hline
 Cycle & 100 & 200 & 1 & 0.26 & 0.64 & 0.75 & 1 & 0.99 & 0.99 & 1 & 1 & 1 & 1 & 1 \\
 Cycle & 200 & 400 & 1 & 0.23 & 0.61 & 0.69 & 1 & 1 & 1 & 1 & 1 & 1 & 1 & 1 \\
 AR(1)& 100 & 200 & 0.99 & 0.26 & 0.65 & 0.75 & 0.65 & 0.99 & 0.99 & 0.99 & 0.99 & 0.99 & 0.99 & 0.99 \\
 AR(1)& 200 & 400 & 1 & 0.23 & 0.62 & 0.70 & 0.99 & 0.99 & 0.99 & 1 & 1 & 1 & 1 & 1 \\
 ER& 100 & 200 & 0.71 & 0.28 & 0.63 & 0.40 & 0.68 & 0.69 & 0.70 & 0.70 & 0.70 & 0.70 & 0.70 & 0.70 \\
 ER& 200 & 400 & 0.80 & 0.27 & 0.68 & 0.75 & 0.79 & 0.78 & 0.78 & 0.79 & 0.79 & 0.79 & 0.79 & 0.79 \\
 Hub& 100 & 200 & 0.50 & 0.23 & 0.48 & 0.01 & 0.47 & 0.48 & 0.48 & 0.48 & 0.49 & 0.49 & 0.49 & 0.49 \\
 Hub& 200 & 400 & 0.44 & 0.22 & 0.43 & 0.16 & 0.43 & 0.43 & 0.43 & 0.43 & 0.44 & 0.44 & 0.44 & 0.44 \\
 \hline
\end{tabular}
\end{center}
\caption{\label{ModSel_LowDim.tab}F1-score of conditional independence graph recovery using different model selection methods.  Results are the average over 50 replications when $N > p$.}
\end{table*}

\begin{table*}[!tbh]
\begin{center}
\setlength{\tabcolsep}{4pt}
\renewcommand{\arraystretch}{1.4}
\begin{tabular}{ |c|c|c|c|c|c|c|c|c|c|c|c|c|c|c| } 
 \hline
 \multirow{2}{*}{Type} & \multirow{2}{*}{$p$} & \multirow{2}{*}{$N$} & \multirow{2}{*}{Optimum} & \multicolumn{3}{c|}{Benchmark methods} & \multirow{2}{*}{IID} & \multicolumn{3}{c|}{Class 1} & \multicolumn{4}{c|}{Class 2}  \\
 \cline{5-7} \cline{9-15}
 &  & & & CV & BIC & EBIC &  & Triangle & ComNei & 4-node & Degree & Triangle & ComNei & 4-node \\
 \hline
 Cycle & 100 & 50 & 0.89 & 0.28 &  0.26 & 0.74 & 0.81 & 0.89 & 0.89 & 0.89 & 0.89 & 0.89 & 0.89 & 0.89 \\
 Cycle & 200 & 100 & 0.97 & 0.23 & 0.29 & 0.69 & 0.76 & 0.96 & 0.96 & 0.96 & 0.97 & 0.96 & 0.96 & 0.96 \\
 AR(1)& 100 & 50 & 0.89 & 0.27 & 0.25 & 0.74 & 0.81 & 0.89 & 0.89 & 0.89 & 0.89 & 0.89 & 0.89 & 0.89 \\
 AR(1)& 200 & 100 & 0.97 & 0.23 & 0.28 & 0.69 & 0.73 & 0.95 & 0.95 & 0.97 & 0.97 & 0.96 & 0.96 & 0.97 \\
 ER& 100 & 50 & 0.54 & 0.28 & 0.26 & 0.40 & 0.47 & 0.49 & 0.49 & 0.49 & 0.36 & 0.46 & 0.46 & 0.45 \\
 ER& 200 & 100 & 0.66 & 0.29 & 0.34 & 0.63 & 0.63 & 0.64 & 0.64 & 0.64 & 0.58 & 0.63 & 0.63 & 0.63 \\
 Hub& 100 & 50 & 0.32 & 0.18 & 0.17 & 0.02 & 0.25 & 0.15 & 0.15 & 0.13 & 0.22 & 0.25 & 0.24 & 0.24 \\
 Hub& 200 & 100 & 0.31 & 0.15 & 0.20 & 0.03 & 0.25 & 0.13 & 0.13 & 0.13 & 0.26 & 0.27 & 0.27 & 0.27 \\
 \hline
\end{tabular}
\end{center}
\caption{\label{ModSel_HighDim.tab}F1-score of conditional independence graph recovery using different model selection methods.  Results are the average over 50 replications when $N < p$.}
\end{table*}

The reason for considering smaller range for $N<p$ is the fact that the graphical lasso estimator does not converge for  $\lambda < 0.1$ in our test cases. The values of $\lambda$ are equally spaced apart with the step size of $0.01$. Multivariate Gaussian data was generated with different precision matrix structures that have been frequently used as test cases in the literature \cite{li2010inexact, tan2014learning}:
\begin{itemize}
    \item Cycle structure with $\Omega_{ii} = 1,\ \Omega_{i,i-1} = \Omega_{i-1,i} = 0.5,\ \Omega_{1p} = \Omega_{p1} = 0.4$. 
    
    \item Autoregressive process of order one AR(1) with $\Omega_{ii} = 1,\ \Omega_{i,i-1} = \Omega_{i-1,i} = 0.5$.
    
    \item \ER\ (ER) structure with $\Omega_2 = \Omega_1 + \delta I_p$ where $\Omega_1$ is a matrix with off-diagonal elements taking values randomly chosen from uniform distribution $\mathcal{U}~(0.4,0.8)$ with the probability of $2/p$ and diagonal values set to zero. To keep $\Omega_2$ positive definite, we choose $\delta = \rho +0.05$ where $\rho$ is the absolute value of the minimum eigenvalue of $\Omega_1$. Here $I_p$ is the identity matrix of size $p$.
    
    \item Hub structure with two hubs. First, we create the adjacency matrix  $A$ by setting off-diagonal elements to one with the probability of $0.01$ and zero otherwise. Next, we randomly select two hub nodes and set the elements of the corresponding rows and columns to 1 with the probability of $0.7$ and zero otherwise. After that for each nonzero element $A_{ij}$, we set $A_{ij}$ with a value chosen randomly from uniform distribution $\mathcal{U}~(-0.75,-0.25) \cup (0.25,0.75) $. Then, we set $\Omega_1 = \frac{1}{2}(A+A^T)$. The final precision matrix $\Omega_2$ is obtained by $\Omega_2 = \Omega_1 + \delta I_p$ with the same set-up as \ER\ structure.
\end{itemize}

Table~\ref{ModSel_LowDim.tab} and Table~\ref{ModSel_HighDim.tab} show the results of applying different methods to recover the conditional independence graph by applying  graphical lasso as the estimator for $N > p$ and $N<p$ cases, respectively. To have a ground-truth for comparison, we provide the highest possible value for F1-score on the regularization path and show it in the column with name Optimum.
For benchmark methods, we considered CV, BIC, and EBIC methods.
The results for CV are given for 5-fold CV, and for EBIC method are given for recommended value of $\gamma = 0.5$ \cite{foygel2010extended}. The results for graph-coding methods are obtained by following the steps outlined in Algorithm~\ref{Alg}. The values are the average of 50 Monte Carlo trials.

It can be seen that graph-coding techniques, i.e., columns named with IID, Class 1, and Class 2, outperform other methods in most of cases. Furthermore, F1-score associated with recovered conditional independence graph by these methods is so close to the optimum value on the regularization path. 
Among graph coding methods, we observed that coding with Class 1 and Class 2 coders give better performance than coding with IID coder in most of the cases. In addition, often times coders in Class 2 may offer a slight improvement over their counterparts in Class 1 coders.
These findings confirm the necessity of having more advanced graph coders that reflect more information about the graph in their codelength.

\begin{figure*}[!htbp]
     \centering
     \begin{subfigure}[b]{0.47\textwidth}
         \centering
         \includegraphics[width=3.6 in]{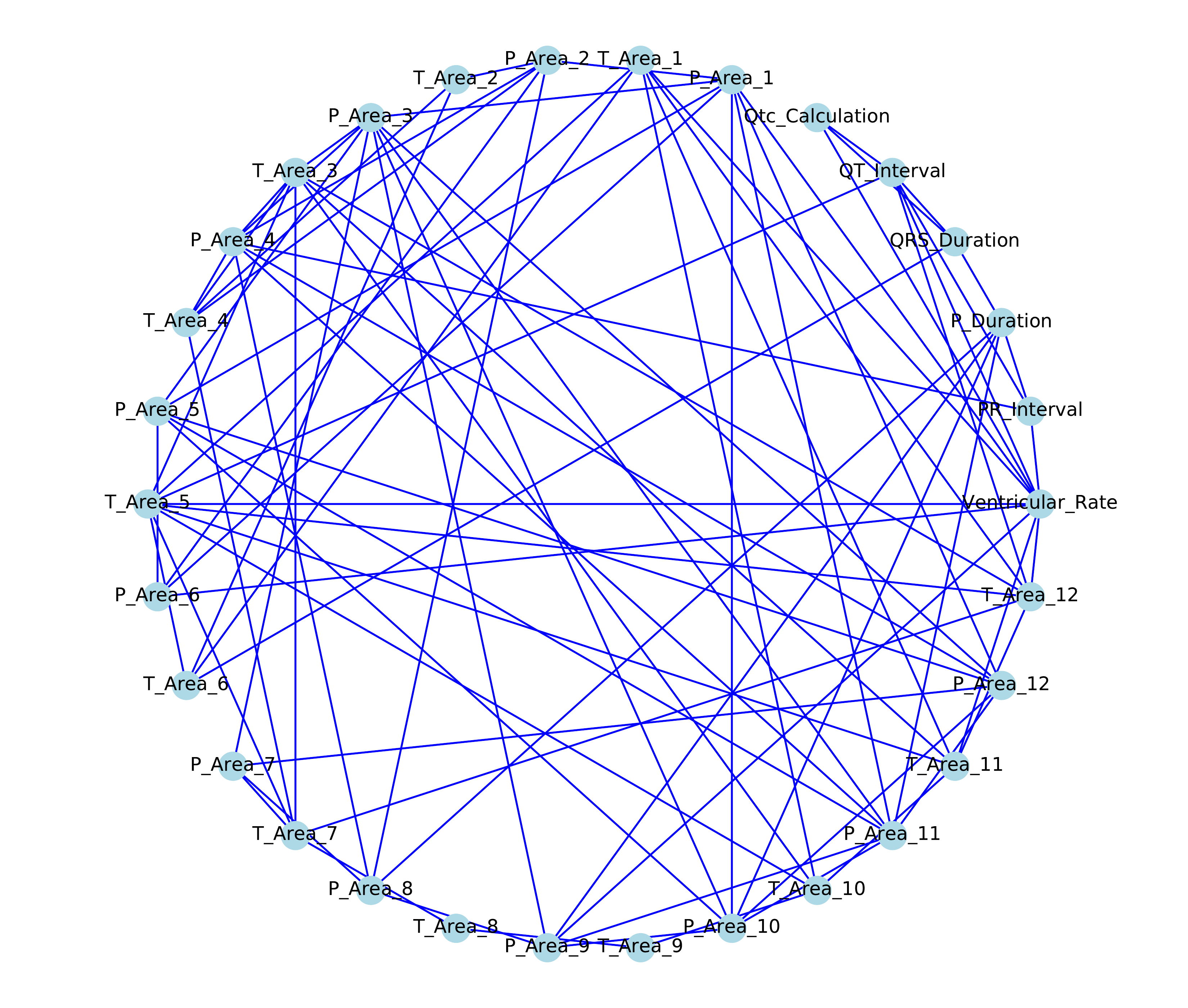}
         \caption{}
     \end{subfigure}
     \hfill
     \begin{subfigure}[b]{0.47\textwidth}
         \centering
         \includegraphics[width=3.6 in]{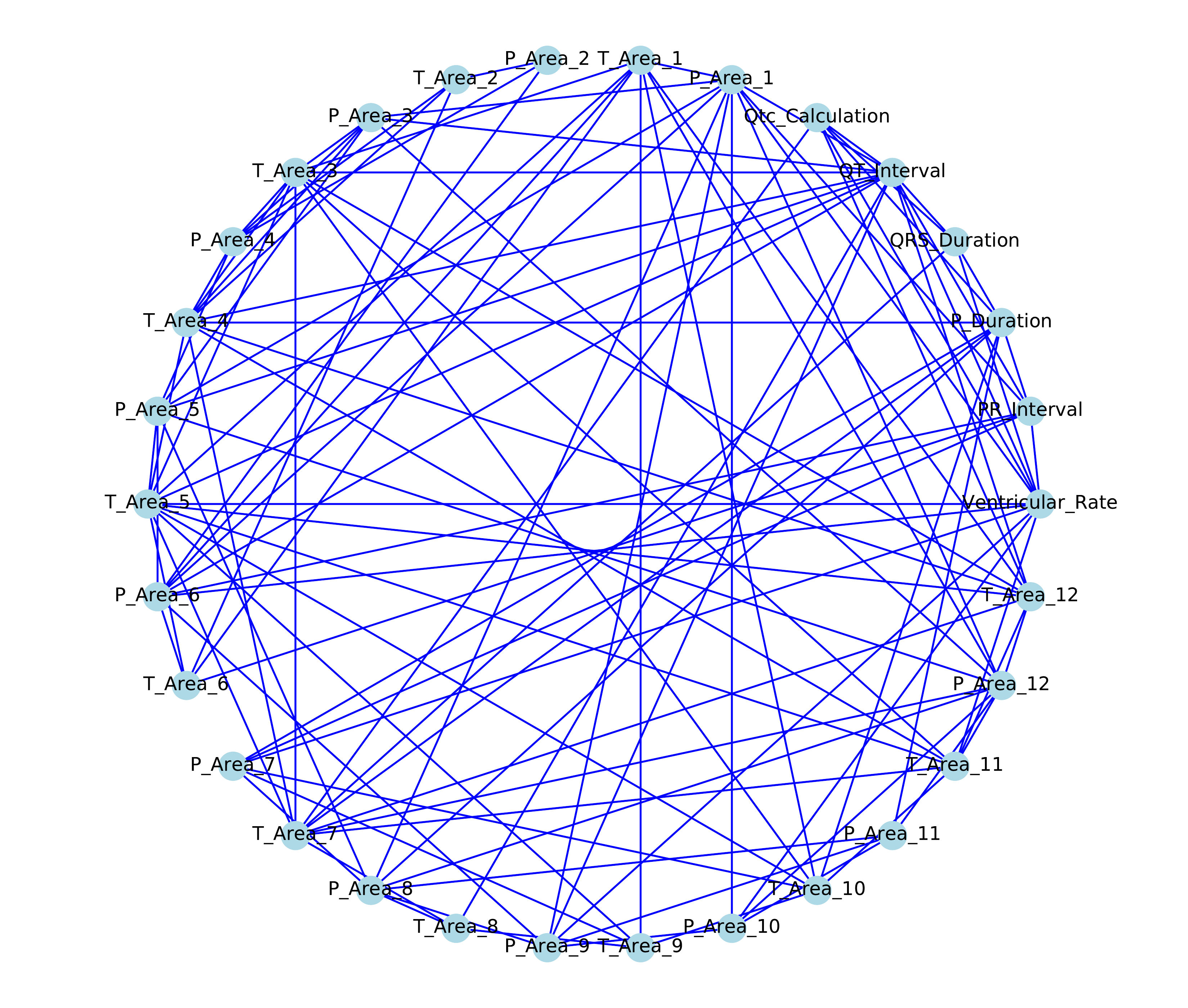}
         \caption{}
     \end{subfigure}
        \caption{\label{GraphModel.fig}Conditional independence graph of (a) healthy subjects, (b) Kawasaki subjects.}
\end{figure*}

We also tested our approach on a real-world dataset. This dataset contains extracted features from 12-lead ECG signals of a group of healthy subjects and a group of subjects with Kawasaki disease. All subjects are of age one to three years. The healthy group has $2492$ samples and the Kawasaki group has $197$ samples.
This dataset was initially processed to select features with empirical distribution close to normal distribution. The reason is that the underlying assumption in Gaussian graphical models is that the data is normally distributed and this is also a key assumption for encoding data under our approach.
After screening all features and selecting those with approximately normal distribution, we ended up with $30$ ECG features. We applied our approach to find the graph model of data for each group of subjects when $\lambda$ takes equally spaced apart values with the step size of $0.01$ within the range $[0.1, 1]$.
The results show that the graph model of healthy subjects differs from the graph model of subjects with Kawasaki disease as given in Figure~\ref{GraphModel.fig}. 
The features with a number from $1$ to $12$ (referring to the lead number) at the end of their name vary across different leads. The rest of features are the same across all leads.
The conditional independence graph given by our approach for each group of subjects is the same across different graph coders. As it can be seen, the conditional independence graph of healthy subjects is sparser with only $87$ edges compared to $115$ edges in the conditional independence graph of Kawasaki subjects. Moreover, these two graphs share $65$ edges.


\section{Conclusion\label{Conclusion.sec}}
In this paper we developed some universal coders based on graph statistics for the compression of unlabeled graphs. We achieved this by first transforming graph structure into a rooted binary tree and showing that the compression of graph structure is equivalent to the compression of its associated rooted binary tree. Then, we used graph statistics to develop two main classes of graph coders. The first class uses graph motifs as local properties and the second class utilizes degree distribution as a global statistics along with graph motifs for coding. We introduced an application of graph coding for model selection in Gaussian graphical models. 

For future work, we will extend this work in two directions. First is to extend graph compression to undirected graphs with attributes where graph topology will be utilized for the compression of structure and attributes. Second is to extend graph coding to directed graphs and using developed coders for data analysis purposes. Similar to undirected graphs, the idea is to transform directed graph into a rooted tree and encode nodal values on the rooted tree.

\bibliographystyle{unsrt}
\bibliography{Graph_Coding_JP}

\end{document}